\documentclass[12pt]{article}
\usepackage{amsmath}
\usepackage{amssymb}
\usepackage{epsfig}

\title{Some comments on the paper ``On the Effects of External Sensory 
Input on Time Dilation'' by Albert Einstein}

\author{ Z.~K.~Silagadze \\
Budker Institute of Nuclear Physics SB RAS and \\
Novosibirsk State University, 630 090, Novosibirsk, Russia }
\date{}

\begin{document}

\maketitle

\begin{abstract}
Einstein's famous 1938 experiment to test relativity of time is plagued 
by too many ambiguities and does not prove anything. Nevertheless, it is 
a landmark experiment at the foundation of the modern theory of time 
perception.
\end{abstract}

\section{Introduction}
Relativity of time, which is at the heart of many special relativistic 
paradoxes, is difficult to understand. {\bf First,} we all are brainwashed
by Newton \cite{1} at our schooldays. Besides, to {\bf learn} relativity
requires a time and effort, and a layman usually lacks both of them.
``The active person lives in the world of phenomena and with it. He does not 
require logical proofs, indeed he often cannot understand them'' \cite{2}.
Why should we try to change such state of affairs and enforce active persons
to understand a bit of modern physics? The reason is simple. Irrationality 
and ignorance is spreading and proliferating worldwide, and such a situation 
is very alarming because  ``a stupid person is more dangerous than a bandit'' 
\cite{3}.

But {\bf how} can we explain relativity of time {\bf to} a layman? Einstein 
himself gave an excellent explanation: ``when a man sits with a pretty girl 
for an hour, it seems like a minute. But let him sit on a hot stove for 
a minute and it's longer than any hour'' \cite{4}.
 
As a genuine physicist, Einstein performed an experiment to prove this 
assertion. Unfortunately, he was a theoretician, not experimenter and his
experiment is plagued by many uncertainties which impede us to {\bf draw}
the above given intuitively appealing conclusion from the experimental data.

\section{Pretty girl and relativity of time}
Einstein's experiment consists of following \cite{4}. He arranged, through 
Charlie Chaplin whom he knew personally, a meeting with Charlie's wife 
Paulette Goddard, the movie star and very pretty woman indeed. When it felt 
to Einstein {\bf as} if a minute had passed during the meeting with the 
radiant and delightful Miss Goddard at the Grand Central Oyster Bar, he 
checked his watch to discover that actually 57 minutes had flied.

Einstein used a chrome waffle iron as a reasonable equivalent of a hot stove,
because the woman who cooked for him had forbidden him to get anywhere near 
the kitchen. He plugged in {\bf the} waffle iron and when it heated up he sat 
on it dressed. He jumped up in less than a second but it felt to him as if a 
{\bf good} one hour had passed. 

At first sight the experiment is very convincing and conclusive. However, as
the {\bf old} wisdom says the first impression is often misleading.

\section{Subjectivity in science is dangerous}
Although the notion of a hot stove is quite well-defined, we cannot say this
about the notion of a pretty girl which is too subjective. And subjectivity in
science is dangerous, as the following story with me does confirm.

I was curious were it the beauty which caused the time dilation effect or the 
proximity of a women and decided to check whether a picture of a beautiful
woman can lead to the same phenomenon, thanks God old {\bf masters} left a lot 
of drawings of beautiful women. I didn't have to go far to find a proper
picture. At the home page of the {\it Southern Cross Review} \cite{5}, 
an on-line magazine where I had found the Einstein paper, there is the 
picture {\it Girl with Guitar} by Anders Zorn who was one of Sweden's foremost 
artists. The girl is quite pretty, albeit naked. There was nobody nearby,
except my wife, to ask whether she was beautiful enough to exclude 
subjectivity. So I called her and asked her opinion. And lo, this cost 
me a black eye {\bf and} the laptop broken to pieces.  

\section{Further problems with the experiment}     
Einstein gave not enough evidence that the pretty girl effect is a real 
effect and not just an illusion. For example, if you sit on a hot stove
in the presence of a pretty girl, the effect evaporates instantly, so
does sometimes the arrival of her husband. It is a grave omission for
Einstein not to investigate these effects and clarify the ambiguity.

\begin{figure}[htb]
     \centerline{\epsfxsize 165mm\epsfbox{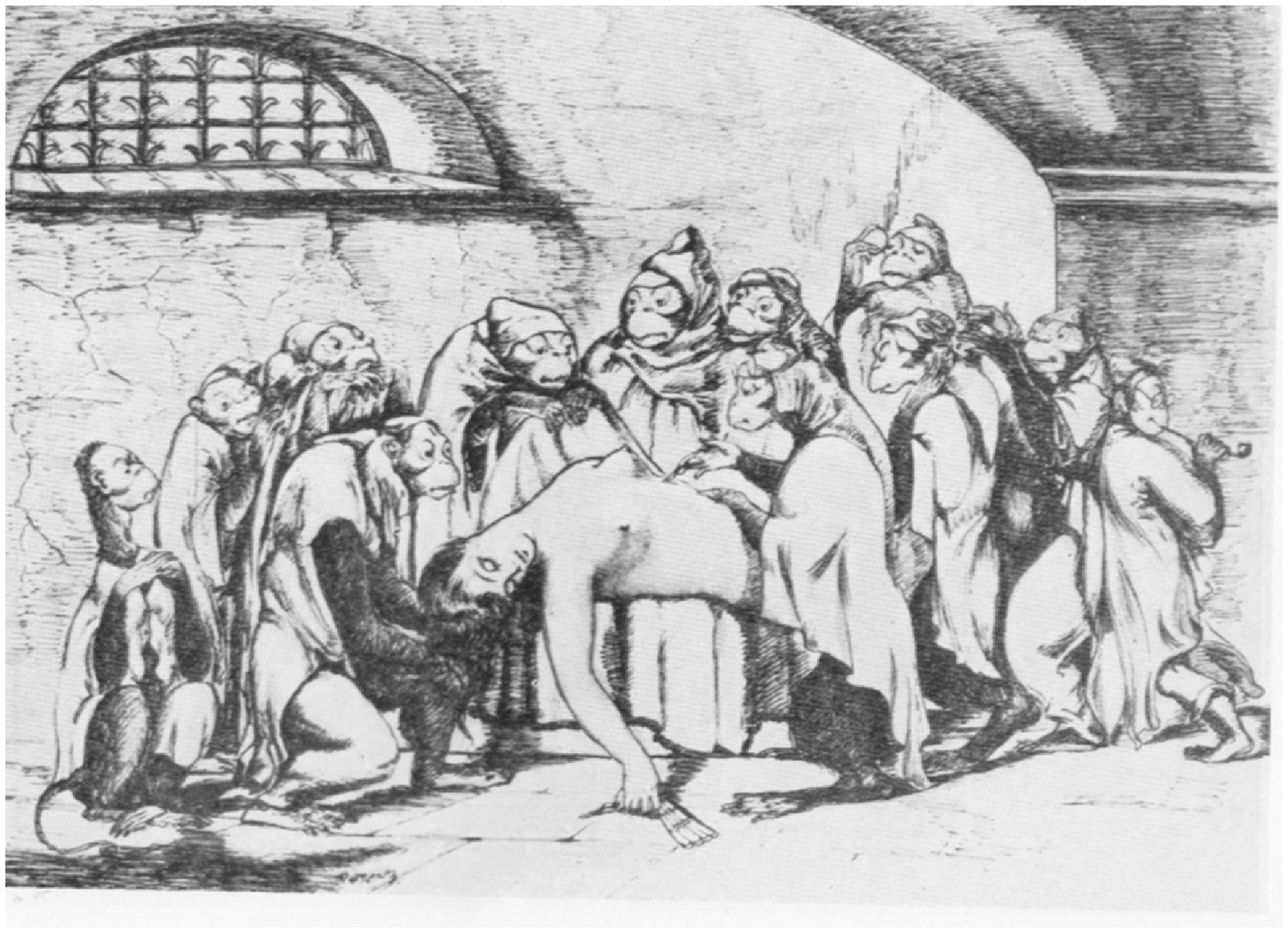}}
\caption{``Quest for the  Secret of Beauty'' by Lado Gudiashvili, 1942.}
\label{Fig1}
\end{figure}

\begin{figure}[htb]
     \centerline{\epsfxsize 150mm\epsfbox{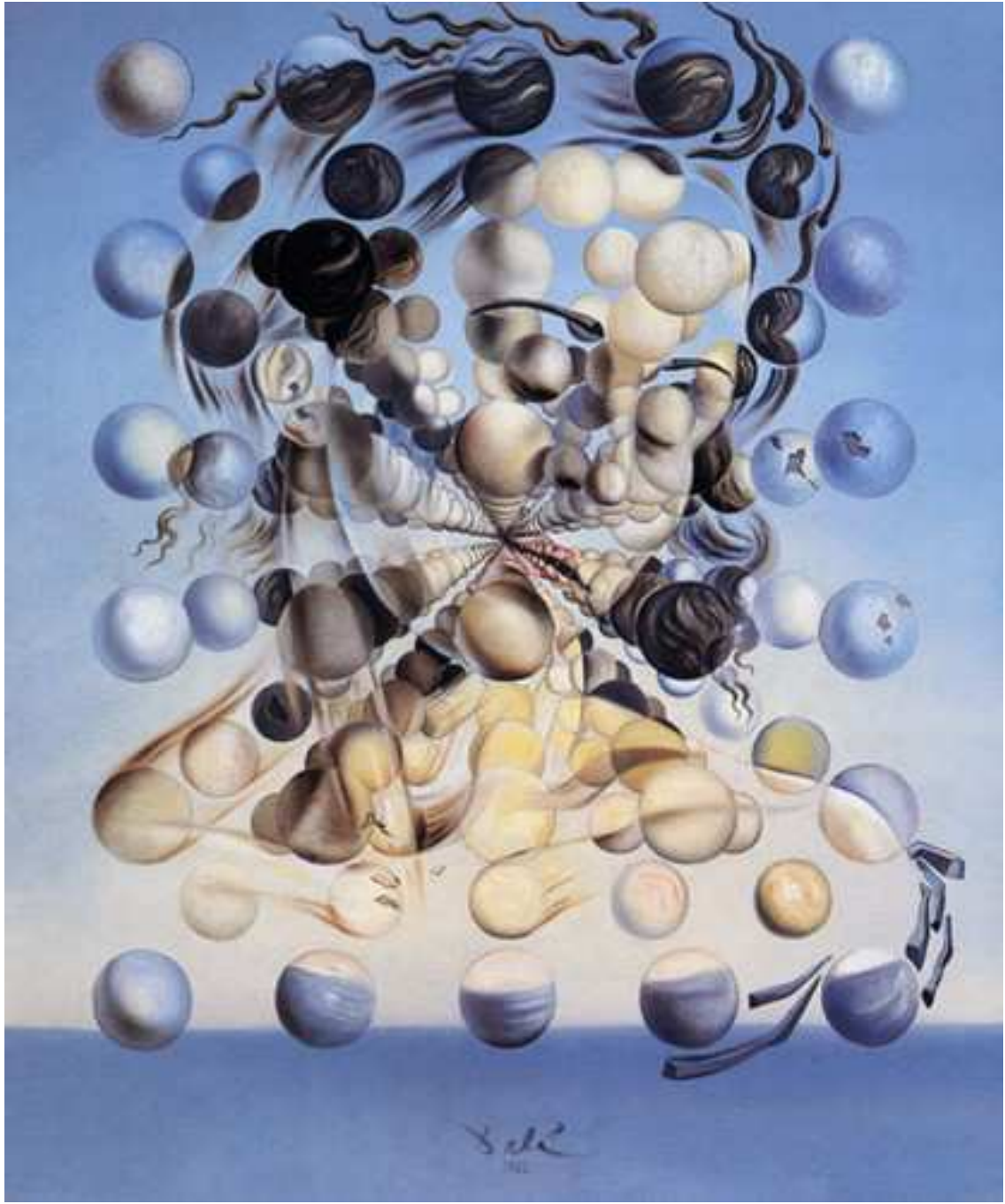}}
\caption{``Galatea of the Spheres'' by {\bf Salvador Dali}, 1952.}
\label{Fig2}
\end{figure}

Neither did he revealed the secret of beauty. Although his fellow scientists
tried hard (see Fig.\ref{Fig1}), this quest for the  secret of beauty is 
still not finished. Most misteriously, the beauty can emerge in quite 
unexpected for scientists way, even from their most beloved thing, a sphere
(see Fig.\ref{Fig2}). 

There is some empirical evidence that the pretty girl effect can change
sign {\bf after} you marry her. That is after years of marital life {\bf you}
can feel like sitting on a hot stove in the presence of your beautiful wife.
This is really very strange phenomenon, more strange than the constancy of
light velocity in different inertial frames. I doubt you {\bf can} explain 
it.

Einstein mentions this mysterious sign change neither in the paper \cite{4} 
nor in his other writings, although it is quite clear that this effect and
the ambiguities mentioned above {\bf do} undermine completely his explanation 
of the time dilation.

Neither does Einstein investigate (theoretically or experimentally) the 
possible influence of a beautiful girl on quantum vacuum, alleged effect
which can be of significant intensity and even dangerous. Namely, it was 
suggested \cite{6} that the presence of a young woman can alter the vacuum 
polarization nearby thus decreasing the molecular bond strengths which by
itself can lead to a spontaneous combustion of materials. Leaving aside 
a danger of spontaneous combustion of a male experimenter, like the notorious 
case of the mysterious death of Krook described by Charles Dickens in 
{\it Bleak house} \cite{7}, particularly disturbing, as far as the Einstein 
experiment is concerned, is the fact that this effect can mimic a hot stove 
and thus blur the time dilation effect.

\section{Einstein and the perception of time}
As we see, Einstein's famous experiment is not without flaws. Nevertheless, 
{\bf whatever} do {\bf you} think, Einstein is a great man, the man of 
tomorrow, if you {\bf want} to know {\bf -} ``while relative time became the 
de-facto view in physics, the relativity of psychological time is still 
a matter for debate'' \cite{8}. However, despite this debate, {\bf everyone} 
agrees (see, for example \cite{8,9}) that Einstein's groundbreaking experiment 
initiated the interesting field of research. This research is far from being 
completed.  ``Although time {\bf is} a concept that attracted and occupied 
the thoughts of a countless number of thinkers and scholars over centuries, 
its true nature still remains wrapped in a shroud of mystery'' \cite{10}. 

\section{Concluding remarks}
There is some on{\bf going} mystery surrounding the Einstein's experiment.
Although Einstein's pretty girl - hot stove explanation of the time 
dilation effect is well documented (he gave it to his secretary, Helen Dukas
\cite{11}), nobody have ever seen the {\it Journal of Exothermic Science and 
Technology} with Einstein's article in it, except Steve Mirsky who reproduced 
this article in \cite{12}. So we are forced {\bf to} repeat Maxim Gorky's 
famous question: ``was there a boy at all?'' \cite{13}. 

Despite all our {\bf respect} for Steve Mirsky, we think what we have here is
an example of unsettled past. It is a common erroneous belief that the past is 
always fixed in every detail. As was conjectured in \cite{14,15}, it may 
happen that the past is actually only partly fixed. At every moment of time 
we have a template of the past that gets incarnated and fixed only under the 
backword influence of ongoing events. By unveiling the Einstein experiment, 
Mirsky initiated a template. However, whether this template turns into 
an undeniable truth, depends on {\bf you.}

\section*{Acknowledgments}
The work is supported by the Ministry of Education and 
Science of the Russian Federation.

\newpage 

\end{document}